\PassOptionsToPackage{dvipsnames,table}{xcolor}
\documentclass[sigconf]{acmart}

\AtBeginDocument{%
  \providecommand\BibTeX{{%
    \normalfont B\kern-0.5em{\scshape i\kern-0.25em b}\kern-0.8em\TeX}}}

\copyrightyear{2024}
\acmYear{2024}
\setcopyright{rightsretained}
\acmConference[ASSETS '24]{The 26th International ACM SIGACCESS Conference on Computers and Accessibility}{October 27--30, 2024}{St. John's, NL, Canada}
\acmBooktitle{The 26th International ACM SIGACCESS Conference on Computers and Accessibility (ASSETS '24), October 27--30, 2024, St. John's, NL, Canada}
\acmDOI{10.1145/3663548.3675612}
\acmISBN{979-8-4007-0677-6/24/10}

\definecolor{lightgray}{gray}{0.9}
\usepackage{balance}       
\usepackage{graphics}      
\usepackage[T1]{fontenc}   
\usepackage{color}
\usepackage{booktabs}
\usepackage{textcomp}
\usepackage{makecell}
\usepackage{xspace}
\usepackage{fancyvrb}
\usepackage{amsthm}
\usepackage{hyperref}
\usepackage{subcaption}
\usepackage{capt-of}
\usepackage{graphicx}
\usepackage{dcolumn} 
\usepackage{tipa}
\usepackage{varwidth}
\usepackage{tabularx}
\usepackage{caption}
\usepackage{color,soul}
\usepackage{geometry}

\usepackage{tikz}
\newcommand*\circled[1]{\tikz[baseline=(char.base)]{
    \node[shape=circle, fill=orange, draw=white, inner sep=1pt] (char) {#1};}}
            
\usepackage{microtype}        
\usepackage{ccicons}          

\newcommand{\eg}{\textit{e.g.}\@\xspace}
\newcommand{\ie}{\textit{i.e.}\@\xspace}
\newcommand{\etal}{\textit{et al. }}

\usepackage{multirow}
\usepackage{enumitem}
\setlist[description]{leftmargin=\parindent,labelindent=\parindent}

\begin{document}

\title{AccessShare: Co-designing Data Access and Sharing with Blind People}

\author{Rie Kamikubo}
\affiliation{%
    \institution{College of Information}
    \institution{University of Maryland, College Park}
    \city{College Park}
    \state{MD}
    \country{USA}
}
\email{rkamikub@umd.edu}

\author{Farnaz Zamiri Zeraati}
\affiliation{%
    \institution{Department of Computer Science}
    \institution{University of Maryland, College Park}
    \streetaddress{4130 Campus Dr}
    \city{College Park}
    \state{MD}
    \country{USA}
}
\email{farnaz@umd.edu}

\author{Kyungjun Lee}
\affiliation{%
    \institution{Department of Computer Science}
    \institution{University of Maryland, College Park}
    \city{College Park}
    \state{MD}
    \country{USA}
}
\email{kyungjun@umd.edu}

\author{Hernisa Kacorri}
\affiliation{%
    \institution{College of Information, UMIACS}
    \institution{University of Maryland, College Park}
    \city{College Park}
    \state{MD}
    \country{USA}
}
\email{hernisa@umd.edu}

\renewcommand{\shortauthors}{Kamikubo et al.}

\begin{abstract}

Blind people are often called to contribute image data to datasets for AI innovation with the hope for future accessibility and inclusion. Yet, the visual inspection of the contributed images is inaccessible. To this day, we lack mechanisms for data inspection and control that are accessible to the blind community. To address this gap, we engage 10 blind participants in a scenario where they wear smartglasses and collect image data using an AI-infused application in their homes. We also engineer a design probe, a novel data access interface called AccessShare, and conduct a co-design study to discuss participants' needs, preferences, and ideas on consent, data inspection, and control. Our findings reveal the impact of interactive informed consent and the complementary role of data inspection systems such as AccessShare in facilitating communication between data stewards and blind data contributors. We discuss how key insights can guide future informed consent and data control to promote inclusive and responsible data practices in AI.

\end{abstract}


\maketitle

\section{Introduction}

Participatory data stewardship~\cite{ada2021participatory} has gained attention as a means to promote ethical and responsible data practices related to collection, sharing, and usage. However, the challenge of ensuring meaningful engagement of diverse stakeholders is pronounced at the intersection of accessibility and artificial intelligence (AI), further raising concerns about the limited agency and control that disabled people may have~\cite{whittaker2019disability}. This is also the case when it comes to their data. With a focus on blind people and the context of computer vision, we see unique accessibility challenges, such as how to support inspection of image data that they have captured and that may contain unintended information~\cite{akter2020uncomfortable} without relying on sight. Despite efforts such as `disability-first dataset' creation~\cite{theodorou2021disability,sharma2023disability}, we still lack mechanisms for providing blind data contributors with access and control over their image data. 

To address this issue, our work aims to explore first what kind of data access and control blind people seek over their image data, as part of our broader investigation into participatory stewardship mechanisms. This is a pressing topic considering how blind people have been early adopters of computer vision technology, sending their photos over servers and assistive applications in exchange for privacy~\cite{gamage2023what,akter2020uncomfortable}. Consequently, they are one of the most represented contributors of publicly-available accessibility datasets~\cite{kamikubo2021sharing}; yet, there is no clear process or consensus among researchers on how to ethically share such datasets. More so, conversations around these data loom large, especially with the tensions involved around benefits and risks---\eg, large multi-modal models are trained on photos taken by sighted people with further widening performance gaps on photos taken by blind users~\cite{cao2022what,massiceti2023explaining}.

In this work, we engineered a larger cross-sectional study designed to foster the involvement of blind participants in challenging real-world data practices from collection to sharing. In this context, we first deployed a teachable object recognizer~\cite{kacorri2017people} in the homes of blind participants, where they captured photos and labels of objects (\ie, study data). These data capturing activities were conducted in real-world settings to bring potential value as AI datasets~\cite{massiceti2021orbit} but came with heightened privacy concerns. Also, the photos were taken via smartglasses that may capture a broader field of view with higher risks of including potentially identifiable information (PII). 
The observations and findings presented in this paper are situated within this larger study, which includes a co-design session to explore blind people's needs and preferences regarding access and sharing of their study data.

Our work encompasses the decisions that 10 blind participants made after the initial data collection and their interactions with their study data via a novel interface called AccessShare. It serves as a design probe~\cite{gaver1999design} to brainstorm potential functionalities and features for an accessible data sharing system. Additionally, our approach facilitates an opportunity for blind participants to access their study data prior to arriving at their decision of whether or not to share them via a public AI dataset. We examined participants' decision-making processes situated in the larger study to understand how decisions were formed, communicated, and evolved over a participatory data stewardship framework.

Our findings revolve around two axes capturing (i) blind participants' data sharing decisions and (ii) requirements for accessible data sharing systems. In the first axis, we found that the majority ($N=7$) made a decision to share their study data. A common rationale, in addition to their motivation behind this decision, was the absence of concerning elements. Participants carefully employed strategies that \eg, avoided capturing personal information and people in the background. Yet, our manual inspection still revealed instances of PII in their photos. We also found that obtaining written data sharing consents via email after the data collection was not effective. All but two participants chose this route. Instead, participants opted for verbal consent during the co-design session. We observed that this is explained in part by a need for an interactive approach to consent. Many participants found the co-design session helpful for guiding their decisions \eg, for better understanding study data and exploring options such as partial sharing.

Findings from the second axis surfaced the complementary nature of data access systems such as AccessShare, in supporting communication with sighted people (\eg, data stewards and family members). Yet, a deeper dive into participants' responses revealed a delicate balance between asking for sighted help and self-reliance when making their decision. While some expressed the need for more reliable descriptors to independently inspect the data, one participant opted for a joint decision making approach with a family member to both inspect the data and share agency with a bystander. The contrasting viewpoints on independence and interdependence prompted considerations for incorporating multiuser functionalities in data access systems. Overall, as a design probe, AccessShare was effective for ideating approaches to enhance data inspection and control and support communication with data stewards.

Our contributions include: (1) empirical findings from a real-world participatory data stewarding process with blind participants, reflecting on decisions formed and how they were communicated; (2) implementation of AccessShare as a probe for co-designing accessible data sharing systems that facilitate control among blind data contributors; and (3) methods for obtaining consent through different means, offering insights into challenges and opportunities for meaningful informed consent for blind data contributors. By illustrating the active participation of blind people across our data stewarding process, this paper can inform future work in involving underrepresented communities in discussions around informed data sharing consents and meaningful data inspection and control.

\section{Background}

We first outline the landscape of data ecosystems and policies in AI, with a focus on individuals' control over their data. We then review prior work on participation frameworks for increasing agency and discuss the value of human-centered research on AI datasets for accessibility. We build upon recent ``disability-first'' data collection efforts, which have highlighted the benefits of engaging blind people in contributing to datasets for AI innovation~\cite{kacorri2017teachable, theodorou2021disability, sharma2023disability} and some of the motivational factors for data contributors~\cite{kamikubo2021sharing}.

\subsection{Prospects for Data Control in AI}

Data in the digital age is a key input and source of value, often referred to as\textit{``the new oil''}~\cite{humby2006data}, a trend that continues with generative AI. A counter-trend is the rising tension to address ethical and privacy concerns surrounding user-generated data~\cite{sebastian2023privacy,khowaja2023chatgpt}. The majority of generative AI applications have been pre-trained on large collections of text data sourced from the internet, including user-generated content on websites and social media platforms~\cite{brown2020language}. Many of these applications also utilize the history of interactions with users to refine and update their models to improve the performance over time~\cite{open2023data}. As data continuously fuels AI, discussions about policy and technical interventions to grant users better control of their data are on the rise~\cite{lacapra2023data,hamza2023chat}. 

Since the launch of the European General Data Protection Regulation (GDPR) in 2018, many users were promised rights related to data protection, including the right to access, erase, and request corrections on their data, and the right to restrict processing~\cite{wolters2018control}. Yet, even when these rights are granted, practical challenges remain, with users struggling to understand, use, and control their data~\cite{bowyer2022human,alizadeh2019gdpr}. Additional challenges arise from the technical and systemic complexity of removing individual data from models once they have been trained~\cite{burgess2023how, basu2024mechanistic}. Tracking the origin and usage of individual data points throughout the AI lifecycle (\eg, training, validation, and deployment stages) can also be challenging, partly due to companies often lacking a full understanding of what has gone into their models~\cite{hacker2023regulating}. While the recent EU’s AI Act could fill in the gaps at scale, enforcing documentation, record-keeping, transparency, and human oversight~\cite{eu2023artificial}, there are yet no clear mechanisms to allow users to have complete data control. 

One crucial aspect of AI and data regulations is the power balance among stakeholders~\cite{de2006privacy}. Despite the regulatory interventions such as the GDPR, Delacroix and Lawrence~\cite{delacroix2019bottom} point to the power asymmetry between \textit{data subjects}--who have knowingly or unknowingly provided their data to various entities--and \textit{data controllers}, typically organizations that make decisions about what data to collect and how it is processed. With our work, we explore methods and tools that can address the current data gaps and facilitate the agency of individuals whose data are part of the AI ecosystem, with focus in the context of research in \textit{`AI for Accessibility'}.

\subsection{Participation in Data Stewardship}
Participatory data stewardship recognizes the importance of involving stakeholders, communities, and individuals in the decision-making processes related to data that affect them~\cite{ada2021participatory}. Ultimately, it aims to rebalance the asymmetries of power~\cite{kapoor2021nudging}. This approach is gaining some traction in AI research and development~\cite{birhane2022power,kelly2023clearing,seger2023democratising}. 

Broadly speaking, participatory data stewardship models have been explored across various fields, presenting both opportunities and challenges~\cite{hafen2019personal,calzada2020platform,delacroix2019bottom}. Data donation, for instance, has been frequently discussed within the medical space involving individuals who voluntarily contribute their health data to a federated dataset for broader societal and collective benefits (\eg, UK BioBank~\cite{uk2022explore}). The key attribute of this approach is that the dataset is created collaboratively by the data donors whose choices such as opting in/opting out to sharing, serve the basis of data curation~\cite{bietz2019data}. However, critics have pointed out that the datasets exhibit self-selection bias, resulting in a lack of representation from minority groups~\cite{kim2019minority}. Other limitations include unclear terms for data donors, such as whether they can restrict the use of their data~\cite{bietz2019data}. 

Data cooperatives have been introduced to facilitate the development of systems and processes that explicitly allow people to gain control over their data~\cite{tait2021case}. An example is Salus Coop~\cite{wilson2022salus} aiming to grant levels of power and involvement to its members. It has developed a \textit{`common good data license for health research'}, which supports the right to specify conditions for using their personal health data~\cite{data2022salus}. The cooperative approach appeals to a sense of data democracy by encouraging members to set policies and make decisions~\cite{swiss2023coop,driver2022coop}. However, increased active participation in data practices presents challenges like the cost of sustaining the infrastructure~\cite{ada2021exploring}. Moreover, potential data contributors may lack the time and knowledge to participate, creating a `data divide' among those involved~\cite{ada2021data}. This calls for more inclusive mechanisms that enable people to shape and govern their data~\cite{calzada2021data}, supported by real-life case studies illustrating attempted participatory methods and the accompanying challenges~\cite{yang2020methods}. In this paper, we build on this call through a case study with the blind community. 

\subsection{Blind Contributors in Accessibility Datasets}

The field of accessibility is no exception when it comes to the growing need for data, considering the potential to facilitate novel assistive technologies~\cite{kacorri2017teachable,morris2020ai,bragg2019sign}. For instance, photos of everyday objects captured by blind users, the focus of this work, can contribute to building image recognition applications to access visual information~\cite{lee2019hands,zhong2013real,sosa2017hands}. There is a rich body of work related to photos sourced from blind people for AI datasets in this space~\cite{gurari2018vizwiz,lee2019hands,massiceti2021orbit,sharma2023disability,sosa2017hands}. Carefully balancing the benefits and unique risks of accessibility datasets is critical in these efforts~\cite{kamikubo2021sharing}. While datasets have more value when collected in the `wild,' capturing real-world scenarios (\eg, people's homes), this comes with heightened ethical and privacy concerns~\cite{heumann2016privacy,gurari2019vizwiz,akter2020uncomfortable}.  

In response, we see efforts toward conscientious data stewardship practices, emphasizing a ``disability-first'' approach~\cite{theodorou2021disability,sharma2023disability,park2021designing}. To facilitate the involvement of blind people in
the creation of a teachable object recognition dataset, Theodorou \etal~\cite{theodorou2021disability} implemented an interface that supports communication between data stewards and data contributors during data collection; for instance, blind people are notified of data that fail to meet the validation criteria (\eg, quality and privacy) and asked to retake them. However, this prompts unaddressed questions about data assessment and the role of contributors. Despite being most affected by the outcomes, blind people are informed rather than actively involved in validation and decision-making processes, partly due to data inaccessibility. For instance, the inspection and filtering of personally identifiable information are done by the data stewards~\cite{theodorou2021disability,sharma2023disability}, not the blind contributors. Thus, privacy-preserving techniques that are accessible to blind people are attracting attention (\eg, ~\cite{zhang2024designing, xie2024bubblecam}). 

Ensuring ethical data practices remains a challenge, especially in engaging blind people to contribute visual content with their consent while respecting their privacy~\cite{sharma2023disability, stangl2023dump}. We complement prior guidelines for disability-first datasets~\cite{theodorou2021disability,sharma2023disability} by addressing how blind people can access, inspect, and better control their data.

\section{AccessShare: Our design probe}

\begin{figure}[b]
    \includegraphics[width=\linewidth]{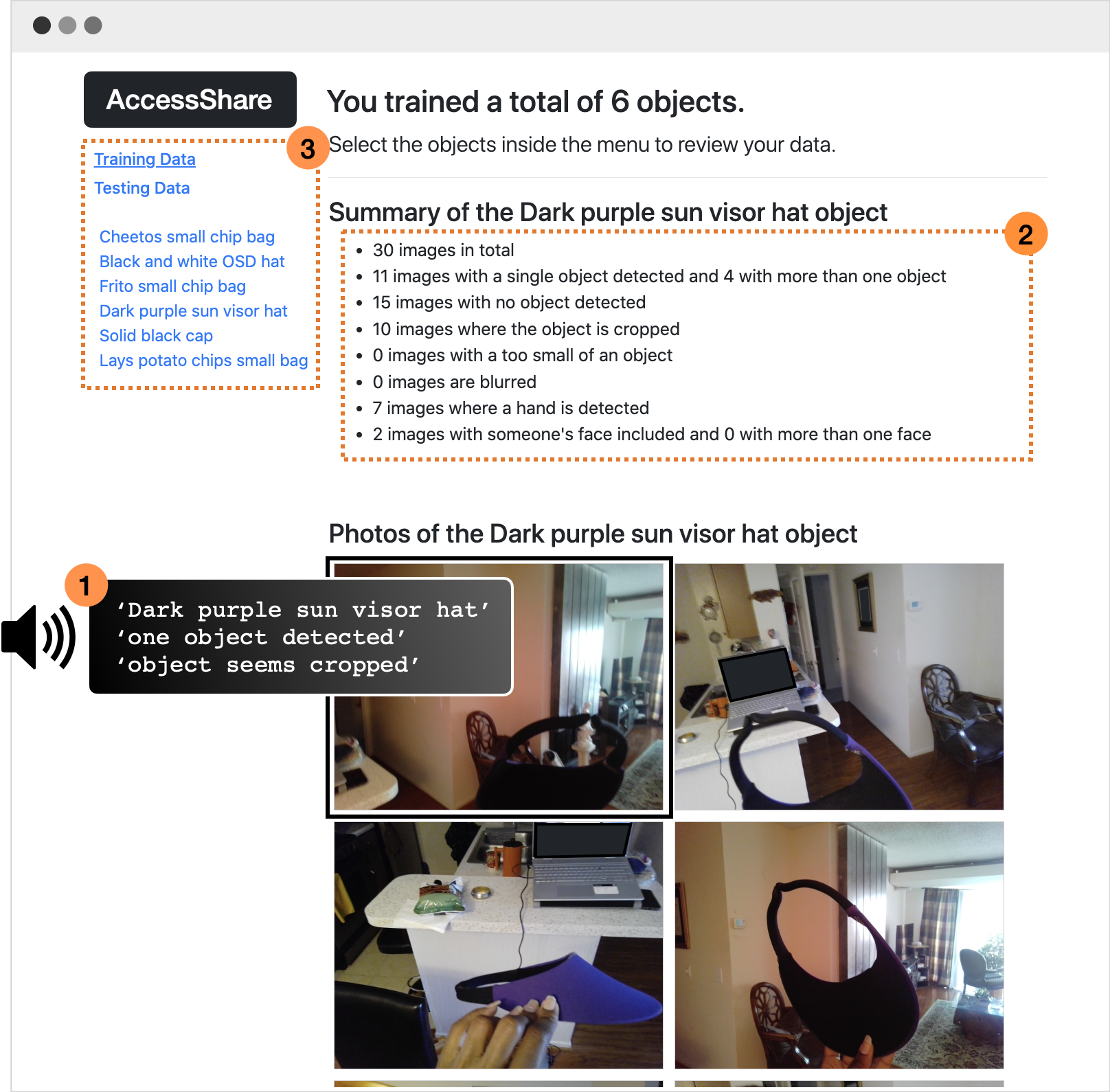}
    \caption{A screenshot of AccessShare, the design probe in our study. Participants review their photos and labels in an HTML interface that contains a navigation menu along a page that provides a summary and a grid view of the photos with alt text based on descriptors for the selected data.}
    \label{fig:interface}
    \Description[Screenshot of the AccessShare web interface]{The left navigation bar, which is highlighted as box number 3, includes options to: 'training data' (selected), 'testing data', and participants' objects such as 'Cheetos small chip bag', 'black and white OSD hat', 'Fritos small chip bag', 'dark purple sun visor hat', 'solid black cup', and 'Lays potato chips small bag'. At the top of the main section of the page there is a heading: 'You trained a total of 6 objects.' followed by: 'select the objects inside the menu to review your data.' The next section is the 'summary of the dark purple sun visor hat object' followed by an itemized summary, which is highlighted as box number 2:
    - 30 images in total
    - 11 images with a single object detected and 4 with more than one object
    - 15 images with no object detected
    - 10 images where the object is cropped
    - 0 images with a too small of an object
    - 0 images are blurred
    - 7 images where a hand is detected
    - 2 images with someone's face included and 0 with more than one face
    The next section includes 'photos of the dark purple sun visor hat object' where there are 4 of the participant's photos visible in a photo grid. The first photo is selected, indicated by a highlighted border. A speaker icon overlays the image, signaling that the app is audibly stating 'dark purple sun visor hat', 'one object detected', 'object seems cropped'. This is highlighted as box number 1.
    }
\end{figure}

To facilitate discussion and ideas for an accessible data-sharing system that can support data control, we develop a design probe~\cite{gaver1999design}. It is called AccessShare, shown in Figure~\ref{fig:interface}.  It incorporates automatically generated descriptors in an HTML webpage that enables blind users to navigate, access, and inspect their data. The probe is closely tied to the data collection task, which in this work is training a teachable object recognizer~\cite{kacorri2017teachable}---users teach the model to recognize objects of their choice by providing photos along labels as training examples. They also take additional photos for testing the personalized model. As seen in previously shared datasets in this context (\eg, ~\cite{sosa2017hands,lee2019hands,massiceti2021orbit}), photos captured by blind people are organized into training and testing, accompanied by files containing object labels. In this section, we detail how the probe is designed for accessing and inspecting these data, which we later discuss with blind participants to brainstorm ideal functionalities and features. 

\begin{figure*}
    \centering
    \includegraphics[width=0.85\textwidth]{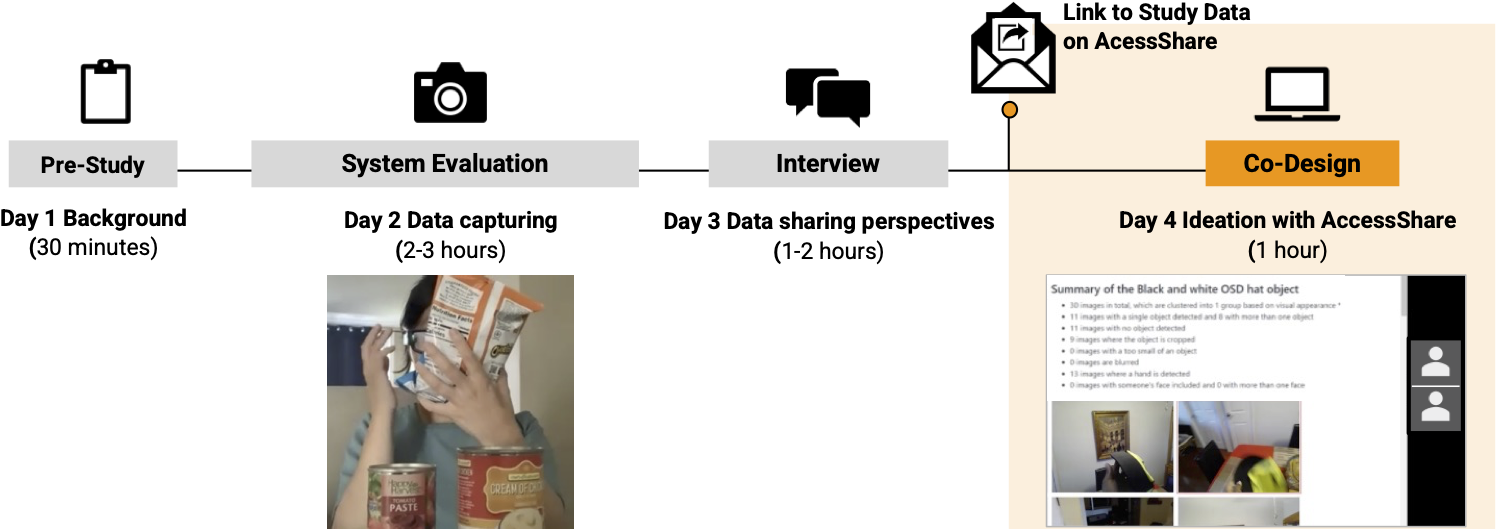}
    \caption{Findings in this paper come from a co-design session within a larger study that includes a background questionnaire, a system evaluation session with data capture, and a semi-structured interview on data sharing. Participants could indicate their data sharing decisions at any time after the interview once receiving an email with a unique link to their data in AccessShare.}
    \label{fig:study_design}
    \Description[This figure shows four different sessions of the entire study.]{From left to right, pre-study interview, system evaluation study, semi-structured interview, and co-design session are shown. During pre-study interview, participants share demographics and experience with technology, and it took on around 30 minutes. In system evaluation study, which took 2 to 3 hours, participants personalize an object recognition application on smartglasses. In semi-structured interview that took 1 to 2 hours, participants share perspectives on sharing data from the study and beyond. In the last one, co-design session, participants design an accessible data inspection and sharing system, and it took on average an hour.}
\end{figure*}

\subsection{ Automatically Generated Descriptors} 

Facilitating a quick turnaround from data collection to data inspection might be necessary to prevent potential data contributors from losing the context needed to reflect on the photos they collected. Thus, instead of manually describing the photos, in our probe we incorporate data descriptors that are automatically generated. We adopt an approach from Hong \etal~\cite{hong2022blind}, which introduces photo- and set-level descriptors for blind users to access fine-grained visual differences across training photos. We find it suitable for our probe in facilitating quick inspection of otherwise similar photos. We repurpose the descriptors to generate summaries and alt-texts for individual photos, enabling blind users to inspect their data. As shown in Figure~\ref{fig:interface}-\circled{1}, photo-level descriptors are embedded as alt-text for individual photos in AccessShare. 

We selected initial descriptors based on two primary factors to assist blind people in determining potential sharing~\cite{kamikubo2023contributing}: i) benefits of sharing datasets to improve real-world recognition tasks and ii) associated risks. In terms of benefits, we considered characteristics that tend to be present in photos taken by blind participants such as ill-framed objects or blurriness~\cite{lee2019revisiting, massiceti2021orbit} that can help future models generalize better for photos with such `image quality'. In terms of risks, we considered privacy concerns~\cite{gurari2019vizwiz}. Specifically, the photo-level descriptors in AccessShare that capture `\textit{image quality}' indicate: the number of objects of interest detected in the photo as `\textit{single object detected},' `\textit{more than one object detected}' or `\textit{no object detected}', photo being `\textit{blur}'; whether the object is `\textit{cropped}'; and whether it appears`\textit{too small}' within the frame. The privacy related photo-level descriptors include:  `\textit{hand presence}' (\eg, for potential identification of tattoos and jewelry) and `\textit{face presence}', as reflected in privacy as well as ethical concerns of blind people~\cite{gurari2019vizwiz, kamikubo2023contributing}. As shown in Figure~\ref{fig:interface}-\circled{2}, set-level descriptors in AccessShare are merely an aggregate of the photo-level descriptors including summary information on the prevalence of those characteristics. Unlike Hong \etal~\cite{hong2022blind}, we do not include detailed set-level descriptors for variation; they were designed to help blind users assess their training strategies to improve their personalized models, a task beyond data inspection for sharing. Instead, the goal of the summary in AccessShare is to help blind users quickly grasp the contents of each set before interacting with individual photos.

\subsection{Interface Design and Implementation} 
We extracted the descriptors using the publicly available code from Hong \etal~\cite{hong2022blind} and stored them in a JSON format. We then used the Jinja web templating engine~\cite{ronacher2008jinja2}, to generate the accessible HTML interface with alt-text for photos derived from the descriptors. We designed a navigation menu reflecting the file structure of photos during the data collection, which follows a two-level hierarchy. The primary menu contains options for `\textit{Training Data}' and `\textit{Testing Data}'. Each opens a secondary menu that provides links to the specific content related to the photos. We named these links according to the object names that users would add as labels during training (see Figure~\ref{fig:interface}-\circled{3}). One critical decision related to split in the file structure was that testing photos are assigned to objects based on the recognition output of a user's model,  that is error prone; thus, they could be misclassified. While these recognition outputs in test scenarios are important for error analysis when shared~\cite{lee2019hands}, a thorough (manual) review is required to correct the labels of testing photos. Since we expect the same challenge in the real world, where only recognition labels are available in testing~\cite{hong2020crowdsourcing}, we used the recognition labels to organize testing photos and explore potential improvements and concerns with blind people.

\section{Methods}

We take a participatory approach to explore what kind of data access and control blind people seek over their data. The ultimate goal is to inform future data collection and sharing frameworks so that they can better align the decision-making process with the needs and expectations of their blind data contributors. As shown in Figure~\ref{fig:study_design}, our co-design session took place within a larger cross-sectional study that involved a series of activities engineered to capture a real-world data stewarding scenario. 
Specifically, we engaged 10 blind participants who were situated in the context of contributing to an AI dataset after evaluating an assistive technology---participants evaluated an object recognition app deployed on smartglasses in their homes and later decided whether they wanted the research team to share their study data via a public AI dataset. To help them decide whether to share or not, we implemented AccessShare, a design probe for inspecting one's study data and invited them to a follow up session where we could design together ways for blind data contributors to access and control their data. 

To provide the context of the larger study, which spanned multiple days, we started with a 30-minute long Zoom call to capture participant demographics, attitudes, and experience with technology. A day or two later, participants joined the system evaluation from their homes and remotely connected with the researchers to perform a series of data capturing activities (described in Section~\ref{sec:systemeval}). Typically within a week after, participants joined a remote semi-structured interview to reflect on their motivations and concerns relating to sharing their study data.
At the end of the interview, participants received an email from the research team, the data stewards, to indicate their decision on whether to share their study data (\ie, anonymized photos and labels) via a public AI dataset. The need to make this decision was communicated to participants early on and was included in their consent forms. 
Participants could also opt to join the follow-up co-design, the focus of this paper, typically conducted a few weeks later via Zoom. Some opted to confirm their decision on data sharing during this final session. We provide more details in Section~\ref{sec:codesign} on our co-design that aims to uncover ways for providing data contributors with more control, followed by our analysis approach in Section~\ref{sec:analysis}. The ongoing analysis from the evaluation of the assistive technology is beyond this paper's scope. Findings from the semi-structured interview are available in~\cite{kamikubo2023contributing}.

\subsection{Recruitment and Participants}
The study protocol was approved by our institution's review board (IRB \#1822836-1). We recruited participants through an existing mailing list in our lab, by submitting a \href{https://nfb.org/research_participant_solicitation_request_form}{request form} for participant solicitation to the National Federation of the Blind (NFB), and via word of mouth. The mailing list in our lab includes blind people who participated in our previous studies and consented to future contact. Members of the list and NFB contacts received our call for participation and consent form via email, and some shared it within their networks. Potential participants emailed their consent to a researcher, confirming they were 18 or older and blind, as required for the study. Participants consented to the multi-day study but were informed they could withdraw anytime with compensation.  

A total of 13 participants were invited to begin via a Zoom/phone call, where we collected demographic information, including age, gender, education, and occupation, as well as technology experience relevant to the system evaluation. After the semi-structured interview, participants received an email with a link to their data via AccessShare to indicate whether to share their data or continue with the co-design; 10 participants opted to participate in the co-design. At the end of the study, participants were also given the option to specify the demographic information that they preferred not to be made available on publication. Table~\ref{tab:participant} reflects their consent. 

\renewcommand{\tabcolsep}{1pt}
\begin{table}[b]
\caption{Self-reported participant information including vision level, age, gender, education, occupation, and AI familiarity on a 4-point scale: 1 = not familiar at all (have never heard of it), 2 = slightly familiar (have heard of it but don’t know what it does), 3 = somewhat familiar (have a broad understanding of what it is and what it does, 4 = extremely familiar (have extensive knowledge). A dash (-) indicates that the participant did not consent to disclose.}
\label{tab:participant}
\small\renewcommand{\tabcolsep}{1pt}
\centering
\begin{tabular}{ | >{\centering\arraybackslash}m{0.07\linewidth} 
|  >{\centering\arraybackslash}m{0.12\linewidth}
| >{\centering\arraybackslash}m{0.07\linewidth} 
| >{\centering\arraybackslash}m{0.12\linewidth} 
| >{\centering\arraybackslash}m{0.16\linewidth} 
| >{\centering\arraybackslash}m{0.17\linewidth} 
| >{\centering\arraybackslash}m{0.2\linewidth} |}
	\hline
PID & Vision & Age & Gender & Education & Occupation & AI Familiarity  \\
\hline
\rowcolor{lightgray}
P1 & totally blind & - & - & -& - & somewhat  \\
P2 & totally blind & 72 & woman & bachelor & retired & extremely  \\
\rowcolor{lightgray}
P3 & totally blind & 67 & man & master & retired & slightly  \\
P4 & totally blind & 48 & man & master & social worker & somewhat  \\
\rowcolor{lightgray}
P5 & legally blind & 32 & woman & bachelor & operations & somewhat  \\
P6 & totally blind & 67 & woman & master & retired & somewhat \\
\rowcolor{lightgray}
P7 & totally blind & - & - & - & IT solution architect & somewhat  \\
P8 & legally blind & 52 & woman & master & application developer & somewhat  \\
\rowcolor{lightgray}
P9 & legally blind & 71 & man & post-bachelor & retired & somewhat \\ 
P10 & totally blind & 67 & man & some college & computer technician & extremely  \\
\hline
\end{tabular}
\end{table}

Of 10 participants who joined the co-design, seven were totally blind and three were legally blind. On average, blind participants were 53.46 years old (STD=14.94). Five self-identified as women and five as men. Participants were compensated \$15/hour, with an average of 2.76 hours (STD=0.51) spent up to the system evaluation (including the opening questionnaires), 1.68 hours (STD=0.36) for the interview, and 1.09 hours (STD=0.13) for the co-design. 

\subsection{Data Collection in Evaluation}
\label{sec:systemeval}

Data collection activities were conducted in our participants' homes, where they evaluated a teachable object recognition app deployed on smartglasses. We aimed to explore realistic concerns beyond the lab settings (\eg, privacy impacted by the environment~\cite{heumann2016privacy}) and data contributions grounded in real-world applications. We asked participants to find a comfortable spot at home to set up the laptop for the Zoom call and interact with the stimuli objects while wearing smartglasses with an embedded camera. We then provided 2 practice objects to help learn how to take photos and provide labels (\ie, object names) using the smartglasses. 

Participants used the touchpad on the smartglasses to navigate the menu and trigger the photo taking and labeling functions, all supported via text-to-speech. They also used voice commands to input, correct, and confirm the object label. Once familiar with the system, participants were asked to complete data capture activities that involved taking multiple photos per object for a total of 6 objects and providing associated labels for training and evaluating a computer vision model. Half of these objects were stimuli provided by the research team engineered to be visually distinct but nearly-identical by touch (\ie different bags of snacks). They were fixed across all participants. The rest of the objects were up to the participant; they could choose anything in their home. Typically, participants opted for products that were similar to the stimuli, such as different canned goods, as shown in Figure~\ref{fig:study_design}.
Participants answered questions related to their experience during the data collection and at the end. 

Each participant generated on average 222 photos (STD=59.9) across the 6 object labels. Both the photos and the labels collected during these activities were referred to the participants as \textit{``your study data''} throughout the communication with the research team; instead, demographic information was referred to as \textit{``metadata''}.

\subsection{Data Access and Sharing in Co-design} 
\label{sec:codesign}
The process for the co-design started right after the semi-structured interview, where blind participants received an email from the research team to indicate their decision on whether to share their study data. The email also attached a unique link to AccessShare, to access their own set of photos organized by labels collected during the evaluation study. They were encouraged to open it on their devices prior to making their decision and were invited to join the research team in a follow up session for designing together ways for blind data contributors to access and control their data. 

The main goals of the co-design session are: to understand the data sharing decisions and the factors that shaped them; explore accessible ways to share study data with blind participants; and gain insights into meaningful descriptors for accessible data inspection.  The co-design activities took place through a Zoom video call and consisted of two parts.

\subsubsection{Part 1: Hands-on interface exploration.} We first asked our participants to open AccessShare on their devices to have an opportunity to explore the interface with the researcher. Some chose to use two different devices, one for the call (\eg, smartphone) and the other to use AccessShare (\eg, laptop), referring to the accessibility challenges of navigating between two windows within the same device. We assigned 3 sequential mini-tasks during this exploration. First, they were tasked with finding a specific object label on the navigation menu. Then they would read a summary content for that object label. Last, they would describe the first three photos on the page after accessing their alt text. 

\subsubsection{Part 2: Discussion with guiding questions.} We then asked participants to describe and reflect on their experience accessing their data. We had a series of guiding questions to foster our discussion and brainstormed potential improvements and features. We list some of the example questions below, which can be described in four primary themes (see supplementary material): 

    \noindent \textbf{Alternative methods:} \textit{What would be an ideal way for researchers to help you make a decision on whether they can share your study data with others or not? Should it be a link to your data? Should it be a phone call where they walk you through what is collected? Should it be just a question at the end of the study whether it is OK for them to share your data?}
    
    \noindent \textbf{Descriptors:} \textit{Was any particular piece of information found on AccessShare helpful to you in accessing what you wanted to know about your data before making a decision to allow for their sharing? What was missing from AccessShare that you wanted to know about your data before making a decision?}
    
    \noindent \textbf{Trade-offs:} \textit{How much would you rely on the automatically-generated image descriptors for making your decision especially when you know that they can be erroneous? Would you still make a decision? Or would you ask for sighted help?}
    
    \noindent \textbf{Potential features and capabilities:} \textit{Are there any AccessShare features that are making it efficient for you to make this decision? Are there any new features or capabilities you would add to it?}

All study sessions were audio/video recorded via the recording feature in Zoom. We also intended to record AccessShare screen activities both to offer guidance during hands-on exploration and to collect observational data. However, this proved challenging in instances where participants accessed Zoom and AccessShare on separate devices. Instead, we relied on the audio from the screen reader to track some of these screen activities during the session.

\subsection{Analysis}
\label{sec:analysis}
To gain a better understanding of how blind participants sought access and control over their study data and how their decisions were shaped in the process, we triangulated data from a variety of sources: recordings from the co-design session, offline communication of participants' decisions around data sharing, and manual annotation of personally identifiable information (PII) in the photos.

\subsubsection{Participant feedback during co-design.} To analyze the data obtained in our co-design session, all participant comments and responses were transcribed. We performed inductive thematic analysis~\cite{braun2006using}, which began with the researcher (R1), who facilitated all the co-design sessions, assigning initial descriptive (\ie, semantic) codes to the data and collating the codes into categories. The research team (including two other researchers, R2 and R3) then conducted ongoing data meetings to add interpretative codes/notes to the categorized data to broaden the analysis. For example, for a data excerpt with an initial code [\textit{preference toward phone call for real-time feedback}] in the category of ``\textit{decision making modality},'' the team added [\textit{rationale: granular decision on data points}] to the excerpt. The process of adding interpretations was completed in two steps: i) the team discussed the sample of organized data (roughly 20\% of all data excerpts) during the meetings, with R2 leading the discussion for an interpretative-level structure and ii) R3 followed the structure to add interpretations to the remaining data. Once done, R1 proceeded with generating and reviewing initial themes and sub-themes in relation to the descriptive codes and interpretations. In synthesis, the (sub)themes were further refined through team discussions to conceptualize them as unifying concepts~\cite{braun2021one}. 

\subsubsection{Participant communication of decisions.} We report blind participants' decisions related to the sharing of their study data. We map these decisions to the analysis of PII in their study data for better contextualization of the rationale behind the decisions. To enrich the overall analysis of their decision making, we also pay attention to their method of consent (\eg, verbal/written) and its timing (\eg, before or after given access to AccessShare) while participating in our data stewarding process.

\subsubsection{PII annotation of photos.} 
Two researchers reviewed blind participants' photos and manually annotated the areas with PII with a codebook that was informed from prior work~\cite{gurari2019vizwiz, theodorou2021disability}. The codebook had two main categories: ``\textit{object}'' and ``\textit{text}''. Under the ``\textit{object}'' category there were different types of PII like [face], [photo of face], [face reflection], [tattoo], and [pregnancy test]. The ``\textit{text}'' category covered various textual PII, such as [book], [business card], [clothing], [screen], [credit card], [letter], [license plate], [menu], [paper], [newspaper], [prescription], [receipt], [street sign], and [poster]. Both ``\textit{object}'' and ``\textit{text}'' categories included a [suspicious] annotation for cases where it was difficult to confirm private content due to poor image quality or complex scenes. Additionally, the ``\textit{other [name]}'' category accounted for any PII instances that were not captured above.

Consistent with prior work, we sought an annotation tool with the following features: (i) can annotate data offline to uphold privacy and data security;
    (ii) supports remote collaboration across annotators; and
    (iii) allows data export for further analysis.    
After a thorough exploration of the available annotation tools, we selected the VGG annotator~\cite{dutta2019vgg} as our tool of choice due to its alignment with these requirements. Using VGG, two researchers from our team first annotated the participant photos individually (3060 photos each) with the order of photos being randomized.

Using Cohen's Kappa~\cite{mchugh2012interrater}, we found a strong agreement between the researchers' annotations for the presence of ``\textit{objects}'' like [face] ($\kappa$ = 0.89), [photo of face] ($\kappa$ = 0.82), and [face reflection] ($\kappa$ = 0.92). For annotations in the ``\textit{text}'' category like [letter] and [clothing], the researchers achieved a moderate level of agreement ($\kappa$ = 0.73 and 0.73, respectively). There were no agreements on the rest of the PII categories: ``text [screen]'' ($\kappa$ = -0.05), ``text [suspicious]'' ($\kappa$ = -0.04), ``object [suspicious]'' ($\kappa$ = 0.0), and ``other [flag]'' ($\kappa$ = 0.0).  

To finalize annotations, researchers resolved disagreements through discussion and reached consensus together. Disagreements on [face], [photo of face], and [face reflection] (N = 9, 221, and 19, respectively) occurred when faces were cropped or partially visible. Often this involved the researchers' faces being shown on the nearby laptop during the Zoom call. Consistent with prior work~\cite{gurari2019vizwiz}, researchers agreed to label partial faces as PII.
Disagreements regarding [letter] and [clothing] (N = 5 and 5, respectively) were initially missed by one of the researchers. Additionally, a new category, ``\textit{other [flag]}'' (N = 92) was introduced by one of the researchers. The rationale behind this addition stemmed from the recognition that a flag could potentially reveal someone's nationality, and some individuals might be hesitant to disclose such information~\cite{kamikubo2023contributing}. Disagreements on [screen] (N = 466) were common and captured the study laptop placed nearby.  When present in a photo, one researcher annotated only the PII inside the screen, such as [photo of a face]. However, the other annotated the entire screen. To mitigate the risk of overlooking any other potential PII content within the screen, in the final annotations, the entire screen was included. The most common disagreement arose in cases where it was difficult to confirm private content due to poor image quality or complex scenes annotated as ``object [suspicious]'' (N = 373) and ``text [suspicious]'' (N = 390). ``Object [suspicious]'' mostly involved photos containing blurred photos of faces. One of the researchers annotated these photos under the assumption that given the similarity among participants' photos, the presence of a face could be inferred from the context.  Researchers resolved disagreements by annotating these instances as ``object [photo of face]''. ``Text [suspicious]'' mostly involved framed papers on the wall or brochures in the background. These instances were jointly annotated as `` text [miscellaneous paper]''.

\section{Findings on to Share or Not to Share}
We explore participants' decisions around the sharing of their data, why they arrived at those decisions, when they were formed, and how they were communicated. We also provide data on PII across participants' photos for better contextualization of the findings with a heatmap shown in Figure~\ref{fig:PII_heatmap}. We then expand on the preferred modalities for participants to form and communicate decisions.

\subsection{What Were the Decisions, Their Rationale, and Connection to PII}

\begin{table}
    \caption{Final decisions and rationale on sharing study data.} 
    \label{tab:decisions}
    \small
    \begin{tabular}{|  c  | @{\hspace{4pt}} p{7em} | @{\hspace{4pt}} p{20em}  @{\hspace{5pt}}|} 
    \hline
         \textbf{PID} & \textbf{Final Decision}& \textbf{Rationale}\\
    \hline
         P1&  not share& found information on data reuse purpose lacking\\
    \hline       
         P2&  share& was motivated to contribute to research\\
    \hline        
         P3&  share > \newline partial share& was not aware of concerning elements in photos > \newline
            thought of an option to remove photos that \nobreak included faces of others\\
    \hline         
         P4&  share& was comfortable after reviewing photos with sighted help;\newline
            found no concerning elements in photos\\
    \hline
         P5&  share& captured photos with privacy considerations;\newline
            found no concerning elements in photos\\
    \hline
         P6&  not share& became aware of privacy concerns\\
    \hline
         P7&  not share& found information to review the data lacking and not reliable\\
    \hline
         P8&  share > \newline partial share& captured photos with privacy considerations;\newline
            found no concerning elements in photos >\newline
            thought of an option to remove photos that \nobreak included faces of others\\
    \hline
        P9& share&was motivated to contribute to research\\
    \hline   
        P10&  partial share & was motivated to contribute to research;\newline became aware of not ``good'' photos\\
    \hline
    \end{tabular}

\end{table}

\begin{figure}
    \includegraphics[width=\linewidth]{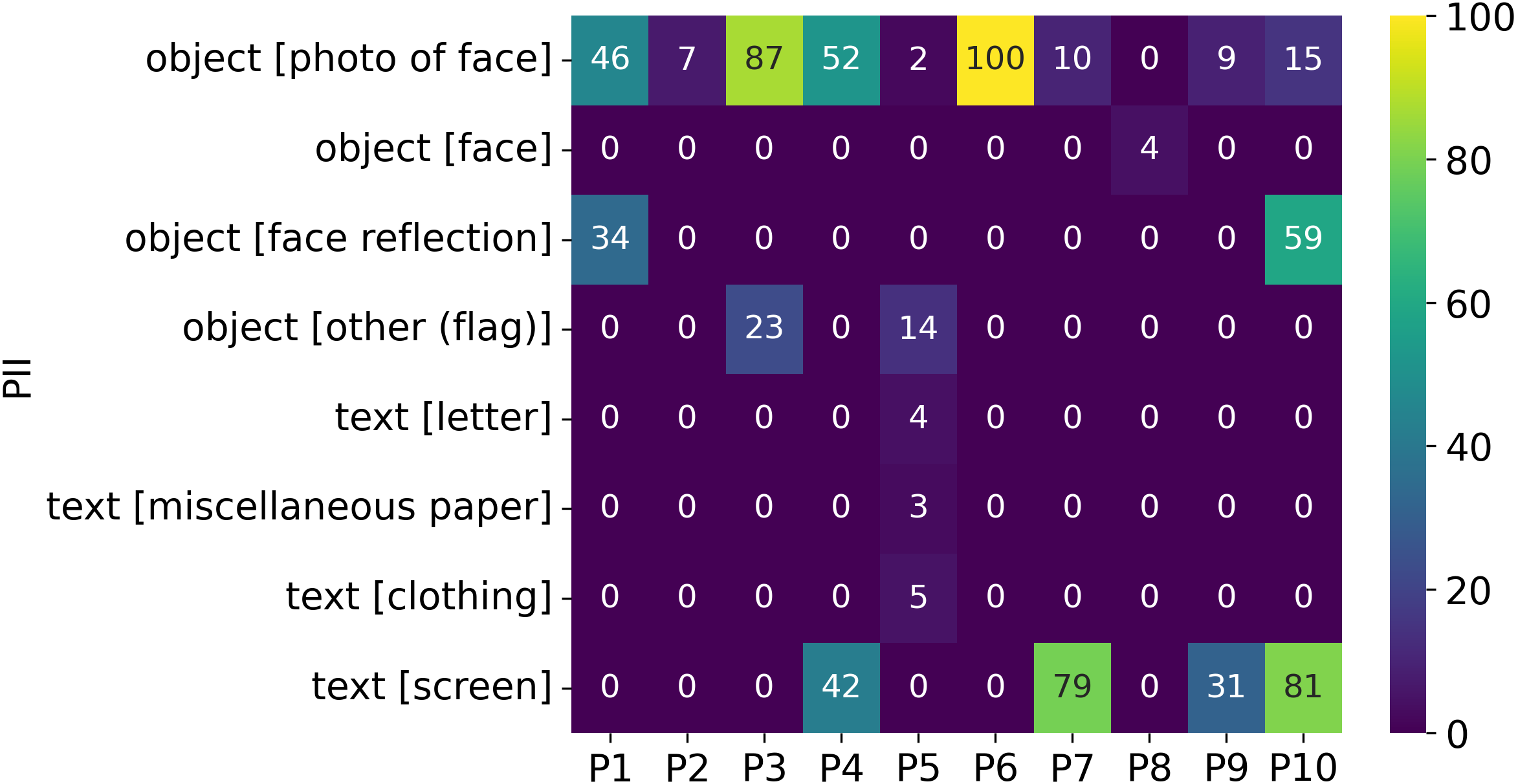}
    \caption{Percentage of PII type across participants' photos.}
    \label{fig:PII_heatmap}
    \Description[A heatmap of PII percentages organized in a 8 by 10 table with rows being the PII labels and the columns the participants.]{
PII: P1, P2, P3, P4, P5, P6, P7, P8, P9, P10 
object [photo of face]: 46.0, 7.0, 87.0, 52.0, 2.0 , 100.0, 10.0, 0.0 , 9.0, 15.0
object [face]: 0.0, 0.0, 0.0, 0.0, 0.0, 0.0, 0.0, 4.0, 0.0, 0.0
object [face reflection]: 34.0, 0.0, 0.0, 0.0, 0.0, 0.0, 0.0, 0.0, 0.0, 59.0
object [other (flag)]: 0.0, 0.0, 23.0, 0.0, 14.0, 0.0, 0.0, 0.0, 0.0, 0.0
text [letter]: 0.0, 0.0, 0.0, 0.0, 4.0, 0.0, 0.0, 0.0, 0.0, 0.0
text [miscellaneous paper]: 0.0, 0.0, 0.0, 0.0, 3.0, 0.0, 0.0, 0.0, 0.0, 0.0
text [clothing]: 0.0, 0.0, 0.0, 0.0, 5.0, 0.0, 0.0, 0.0, 0.0, 0.0
text [screen]: 0.0, 0.0, 0.0, 42.0, 0.0, 0.0, 79.0, 0.0, 31.0, 81.0
}
\end{figure}

As shown in Table~\ref{tab:decisions}, the majority ($N=7$) of our participants made a decision to share their study data including photos of objects and labels. A common rationale for deciding to share was related to the absence of concerning elements in the photos. P3 and P4 came to this conclusion after spending at least an hour reviewing their study data with AccessShare, with P3 doing so independently and P4 with a sighted family member \textit{``to make sure that something did not end up in there that I don't want shared with everybody''} (P4). Surprisingly, both P3 and P4 had many photos (87\% and 52\% respectively) in which pictures with recognizable faces (\eg, found in a bookshelf in the background) were visible, as shown in Figure~\ref{fig:PII_heatmap}. After discussing these PII with the researcher, only P3 changed their initial decision from `share' to `partial share,' omitting the photos with faces.

When talking about their decision to share, others like P5 and P8 referred back to their privacy strategies employed in data collection, reflecting their preparation for data sharing. For example, P5 said, \textit{``I purposefully, when I took the images, made sure that it was pointing towards the carpet or towards the wall. So I know that there's nobody in this photo, so I feel comfortable sharing.''} Indeed, these two participants had the smallest occurrence of PII in their photos. Yet, the few PII included could pose a high privacy risk. For instance, 4\% of P5 photos included a letter capturing both the sender's and receiver's information that given a better image resolution would be directly visible. Similarly, 4\% of P8 photos captured a fully visible face of a family member. Upon discussion of these PII with the researcher, P8 changed their initial decision from `share' to `partial share'.

As an overarching rationale behind data sharing, P5 and P8's cautious but not bulletproof data capturing strategy could also reflect blind participants' underlying motivations for joining the study. Their main goal could be to contribute to research, which they expected to achieve through data, as specifically expressed by P2, P9, and P10. For example, P9 said, \textit{``I knew right away [to share their study data]...I don't want to be an inhibitor to say no, you can't look at my data that would be foolish. Why would I want to participate in the study? I'm here to help you. If you can't learn from my data, then what am I as a human subject?''} We suspect that by \textit{``you''} P9 is referring to our research team in particular instead of sharing more broadly. Among these participants, P2 and P9 only included photo of faces captured from the study laptop screen including the Zoom setup in 7\% and 9\% of their photos, respectively. However, P10 had a greater number of PII, capturing their own face reflection from the mirror in 59\% of their photos.

Looking at those who decided not to share their data, we also found that the most common resistance stemmed from potentially concerning elements in the photos. During the co-design session, despite having the inclination to share initially, P6 ultimately opted out from sharing after discovering that their captured photos contained unintended information in the background. Indeed, all (100\%) of P6’s photos were flagged as including photo of face, which were family pictures captured in the background (\ie, framed on the wall). 
P7, who was initially inclined not to share, did not change their decision. Their rationale revolved around the automatically generated descriptors that were perceived to lack the reliability needed to identify concerning elements. P7 said, \textit{``Based on the fact that there's not enough information for me to review it reliably in the sense of-- the picture descriptions don't have enough...It's no, because I can't make an informed decision.''} Our manual inspection revealed that in the majority (79\%) of the P7’s photos, the study laptop screen was visible but a smaller portion (10\%) included their face in the Zoom call as a small icon. Not surprisingly, the underlying face detector models used for the corresponding AccessShare descriptor failed to detect this small screen icon as a face in most photos where it was present.  
Making an informed decision was also a challenge for P1, whose photos included the faces in the Zoom call captured from the laptop screen (46\%) and often their reflections in the glass table (34\%). Yet, P1 found making an informed decision challenging for a different reason, stressing the lack of information regarding how the study data would be reused by others once shared. 

Interestingly, our participants' willingness to share was also often impacted by the perceived value of their study data. 
While P10 discovered potentially concerning elements (\eg, their own reflection in a mirror), that did not deter them. When they  ultimately consented to partial sharing by removing photos with those elements, their rationale did not revolve around concerns of privacy but around concerns of quality. 
A common theme across participant feedback related to the quality of their data, was referring to them as \textit{``good''} or \textit{``bad''} photos. With the use of descriptors that indicated the presence of faces, hands, ill-framing, blurriness, etc., most of the participants (all except P3, P4, P6, P7) raised their hesitancy to share their \textit{``bad''} photos as a \textit{``a path of interference''} (P9) to the research. For instance, P5 indicated it as more concerning than sharing photos that captured a family member: \textit{``The only thing is that some of the pictures are not great. I hate to share bad photos. Otherwise, I'm okay [with sharing].''} Consequently, AccessShare was often perceived as a useful tool to reflect on their data collection and take \textit{``better''} photos in the future, as referred by P9: \textit{``If there were a session to follow up and for me to take more photos, I would get rid of that clear glass table. I would probably not use my hand as much. I would put the object on the table and center it and not have any other interference. I've learned already what to do better.''} However, the value of datasets is often relative to the needs and contexts of the models that the data support. For instance, real-world photos captured by blind users and sent to object recognition systems may include elements like out-of-frame or blurred objects. Thus, datasets that are shared are most helpful when they match these real-world scenarios where the technology is deployed~\cite{gurari2018vizwiz, lee2019hands, massiceti2021orbit}. Our participants' perspectives on the \textit{'quality'} of their data reveal a unique \textit{`value'} gap between blind data contributors and data stewards who aim to construct datasets representing the blind community.

\subsection{When and How Decisions Were Formed and Communicated}

\begin{figure}[t]
    \centering
    
    \includegraphics[width=\linewidth]{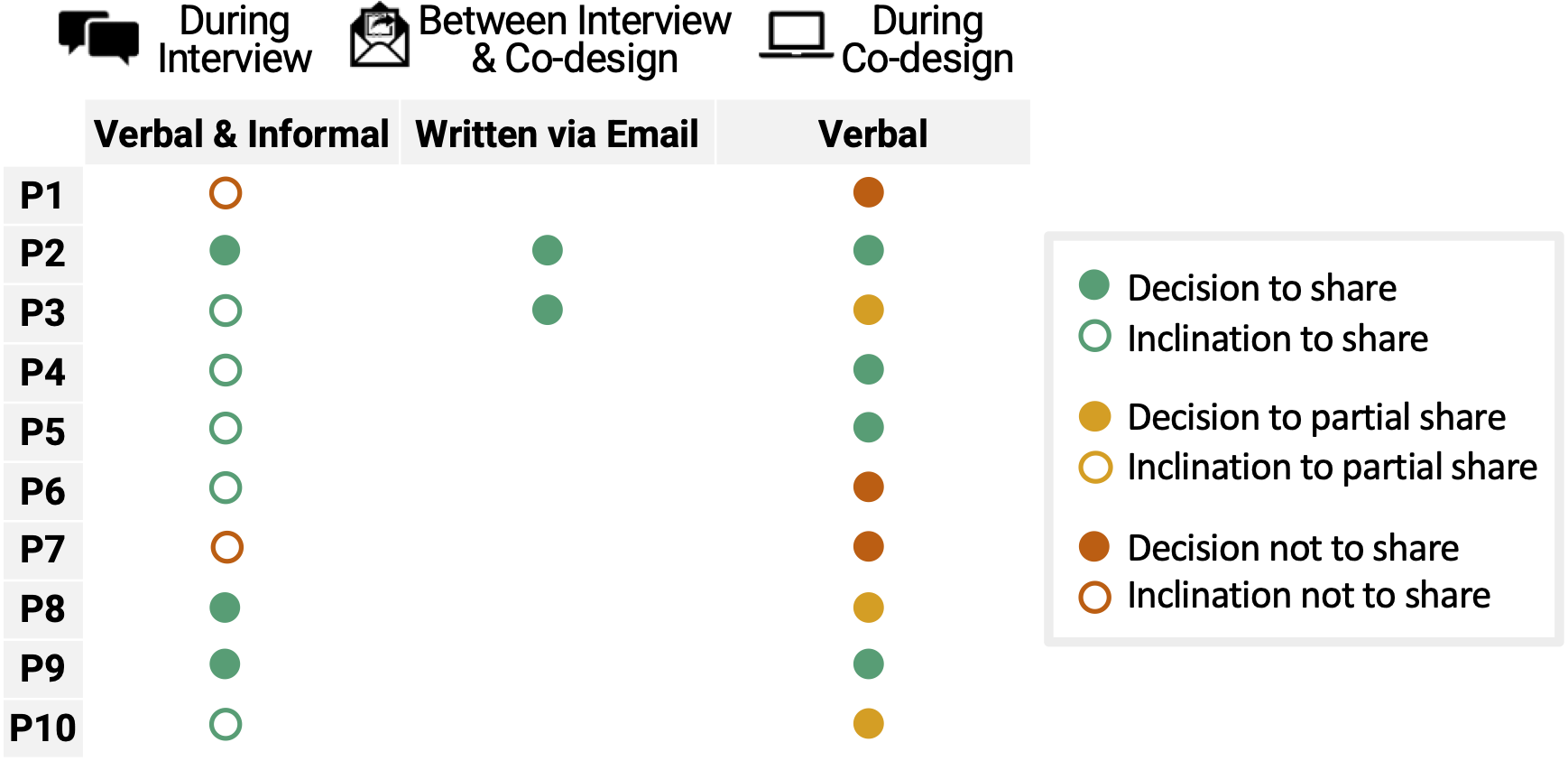}
    \caption{When and how decisions were communicated.}
    \Description[A 10 by 3 table with rows being the participants and columns being when and how the decision to share was communicated.]{
    When and how: During interview (verbal and informal), Between interview and co-design (written via Email), During co-design (verbal)
    P1: inclination not to share, none, decision not to share
    P2: decision to share, decision to share, decision to share
    P3: inclination to share, decision to share, decision to partial share
    P4: inclination to share, none, decision to share
    P5: inclination to share, none, decision to share
    P6: inclination to share, none, decision not to share
    P7: inclination not to share, none, decision not to share
    P8: decision to share, none, decision to partial share
    P9: decision to share, none, decision to share
    P10: inclination to share, none, decision to partial share
    }\vspace*{-2pt}
    \label{fig:decisiontimeline}
\end{figure}

As shown in Table~\ref{fig:decisiontimeline}, when and how participants communicated their decisions differed. Initially, some participants (P2, P8, P9) indicated their decision informally during the interview study, which occurred before they were given access to their study data. After the interview, only two participants (P2, P3) chose to provide written consent by responding to our email; we received their responses within a few days. P2 re-confirmed their decision to share, and P3 confirmed it for the first time after accessing their study data. The majority ($N=8$) of the participants opted for verbal consent only which they conveyed while participating in the co-design session. Even those who ($N=3$) who opted out of the co-design study altogether, did not respond back to the consent email. This indicates that although emails may appear to be an appealing approach for data stewards to obtain written consents, their effectiveness may be limited for blind data contributors, regardless of whether they are sent at the end of the study or between study sessions.

We next look at different factors that may have influenced how participants' decisions were formed. In the interview study, participants were prompted to reflect on the potential benefits and risks associated with the sharing of their study data~\cite{kamikubo2023contributing}.
This facilitated the communication of data sharing perspectives and decisions across participants (specifically indicated by P2, P8, and P9). For example, P8 referred back to the interview: \textit{``During the interview I've already gave my opinion regarding the photos, what kind of photos I'd like to share''}. Our participants' overall inclination to share (P3, P4, P5, P6, P10) or not share (P1, P7) was apparent in this phase.

With the addition of AccessShare, the discussion and reflection on study data became more situated during the co-design session. It helped participants discern their concerns and preferences, as noted by P1 and P5, whose initial inclinations were \textit{``solidified''} (P1) or \textit{``reinforced''} (P5). P1 added, \textit{``It has made me think a good deal about what would make me want to share and what wouldn't.''} Moreover, the interactive nature of the study was deemed helpful in understanding the researcher's intent in forming their decision (P5, P7, P8). As prior work surfaces `trust' toward data stewards as a critical component in data contributors' judgments of concern and risk~\cite{mozersky2020research,kamikubo2023contributing}, P7 articulated the impact of participation on their initial inclination: \textit{``What's changed is that I was able to bring concerns to you and you were willing to listen and hopefully address them. That leads in a bit more positive towards how you might handle it.''}

How decisions were formed might relate to AccessShare, potentially serving as an effective medium to communicate and address concerns with sighted help. Two participants (P6, P10) found the dialogue with the researcher helpful in filling in the role of sighted help to assess the sensitivity of potential PII: \textit{``It's helpful because you [the researcher] clarified some things for me about the face...without really understanding what the data is, it's hard to make decisions about whether you share it or not''} (P6). For a better understanding of their study data, P4 chose to use AccessShare with a sighted family member, indicating how the decision was formed prior to the co-design session: \textit{``I've already done so much as far as reviewing the data and talking [with a family member].''} Additionally, their scanning \textit{``things that we didn't want''} (P4) using AccessShare, might reflect their valueing of the opportunity to address shared privacy concerns~\cite{akter2022shared} and engage in joint decision making.

We could further see how the ideation process during the co-design session impacted the exploration of options related to their study data, particularly in the decisions of two participants who switched (P3, P8). While the photos they captured contained unintended information, such as the experimenters' faces visible on a computer screen in the background, it was not critical for them, as articulated early in the session by P3: \textit{``I'm glad you told me but I'm still inclined to release everything because the information in the photos, there's nothing as far as I know that I would care about.''} Yet, as they went through brainstorming for ideal sharing methods, they started to consider a different option. By the end of the session, they solidified their decision to share partial data, asking the researcher to remove the photos with faces. P3 shared the impact of participation in the co-design in exploring more control: \textit{``I didn't know this was possible. I would like to consent to releasing everything except photos where the person's face. So in that respect, it [participation] did influence me. And I think influenced me for the good.''}  

\vspace{-0.4em}

\subsection{What Were the Decision-Making Modalities}

When asked about their preferred decision-making approaches, participants expressed mixed perspectives on the modalities involved. The primary modality chosen by participants was typically a single question at the end of the study (P2, P3, P5, P10) or a phone call (P8, P9) allowing for verbal exchanges. Written communication like emails was the least preferred option among those wanting to convey granular decisions in real time (P4) and reduce the chance of misunderstandings (P5). Our participants' tendency to provide verbal consent as opposed to written consent could be reflecting this preference. Still, to authorize permission for sharing, one participant (P1) stressed the need for a written consent process via email or a form to communicate data storage and purpose, along with assurances regarding anonymity and privacy.

As we explored these preferred consent methods, participants expressed different opinions regarding the role of AccessShare. While some (P1, P3, P8) perceived data access as supplementary, described as a \textit{``nice option to have''} (P8), some others (P4, P6, P7) considered it a complementary modality for decision making. This preference for data access as complementary (except for P7) seemed to tie closely with accessing sighted help: \textit{``The link to the data was helpful for me, because I had somebody go through the photos with me, just to make sure what we were looking at that was in the background''} (P4). On a different end, P7 preferred a system with enhanced image descriptors and mechanisms that can eliminate the need for sighted help, since \textit{``it undermines the very point of a system like this if we need it [sighted help] in the first place. Otherwise, what's the point?''} 

When asked to expand on the potential use of AccessShare as a communication interface for decisions, participants elaborated on some design features that could reflect their preferences. There was a suggestion for implementing an approval/disapproval button (P4, P8) or a comment feature (P8) on individual photos to communicate partial sharing. Being concerned about transparency related to data reuse purposes once shared, P1 also suggested a way to directly interact with those requesting to access their study data, which they would need to provide \textit{``purpose, the uses, who else would have access''} in order to grant permission.

\section{Findings on Accessible Data Sharing}

We explore participants' perspectives on designing an accessible data sharing application for blind people. We report their experience with AccessShare focused on the current state of the interface and descriptors, as well as their feedback on improvements from our ideation activities to further understand design opportunities.

\subsection{What Was the Experience with AccessShare}

Overall, almost all participants found AccessShare easy for navigating their study data, with its accessible structure and guidance. The only exception was P7, who had some difficulty on a computer compared to the more \textit{``straightforward and streamlined''} mobile interface. Two participants (P7, P9) wished for a distinct homepage introduction as a starting point for navigation. They suggested a clear heading such as \textit{``[participant's name]'s dataset''} (P7) or a short \textit{``human subject summary''} (P9) outlining the objects that they trained and tested.  The majority of the participants (excluding P3 and P4) expressed spending little time on AccessShare, often less than 10 minutes, with P2 and P8 specifically mentioning the perceived ease of the user interface as a contributing factor. 

The group-level summary of the photos by descriptors was found to be the most efficient feature of AccessShare by some (P1, P3, P6). It was seen as more helpful than going through individual photos to flag potential concerns, as noted by P3: \textit{``The summaries were extremely helpful, or helpful than the descriptions of the photos. Well, because for example, the one we just looked at where it said, I think it said four photos, there was a face that (...) that immediately told me something's wrong. In other words, that was sufficient to tell me, yes, something's wrong. And by seeing it in the summary, I didn't really need to know where it shows each individual photo.''} We found that participants often reviewed the summary for each object and only a few or no photos for individual descriptions.

Participants found the ambiguity in the descriptors as one of the major limitations of AccessShare, mostly concerning the presence of the target objects (P1, P3, P5, P6, P7). For example, P7 said, \textit{``It just says one object detected, multiple objects detected, hand detected. But it doesn't really tell me if it detected the object that I had intended it to be.''} Some participants (P3, P5, P6) were further confused by vague descriptions like `\textit{more than one object detected}' without specifying the extra objects. This perspective mirrors a trend seen in prior work~\cite{macleod2017understanding}, where blind participants preferred more descriptive photo information, despite occasional errors, over consistently correct but less detailed information. Others  (P4, P10)  found some value in this partial information; for example, \textit{``It does let you know that there might be more than just what you're trying to take...It provides some useful information. Obviously, that is not good as human eyesight, but there's not really a whole lot that replaces that yet''} (P4). Still, the usefulness of certain descriptors was questioned by some (P2, P4, P6, P7); for example, the `\textit{hand presence}' descriptor was providing already known information, as articulated by P4: \textit{``I knew my hand was in all the photos. So that was kind of already a given.''}

\subsection{What Were the Perspectives on Future Descriptors and Data Access Functionalities}

Ideation with participants revealed opportunities for extended descriptors and data access functionalities. 

\noindent\textbf{Filter \& Sort.}  One of the popular design suggestions was to add a filtering feature, often by descriptors (P3, P5, P6, P8, P10). This idea revolved around the need to inspect the study data that are flagging concerns at an individual photo level, especially photos with `\textit{face presence}.' Additionally, a few participants (P3, P5) suggested a filter by the number of objects detected in the photos to address the confusion brought by the `\textit{more than one object detected}' descriptor for further inspection. Reflecting a similar need, a sorting feature was mentioned (P3) to allow users to inspect from the most problematic---\eg, with `\textit{face presence}'---to less problematic photos. 

\noindent\textbf{Edit \& Delete.} We also identified a user need to refine their collection of labeled photos of objects (P2, P3, P7, P10). Design suggestions varied. For instance, P3 and P7 suggested first a grouping of potentially problematic photos, via the `\textit{face presence}' descriptor (P3) or identification of \textit{``potentially compromising information or personally identifiable objects''} (P7), then verification and correction of classification errors. Another common suggestion was the removal of photos, either those containing PII (P3, P7) or unwanted characteristics such as cropped objects or hand presence (P10). However, this could compete with the need to collect real-world data. Instead, inspired by P2 feedback, such functionalities could be more fine-grained where participants could delete a photo from the `public' collection, meant to be publicly shared, but not from the `study' collection, meant to be shared with the specific research team only. 
    
\noindent\textbf{Search.} When prompted to discuss a potential search feature, a few participants (P3, P7) expanded on its design. For example, P3 mentioned the identification of PII content, with the ability to search by face or text: \textit{``If you're wearing a shirt that says [name of school] or something like that, I'd like to know that. Or who knows, there could be something in a picture that let's just say is inappropriate.''} This perspective aligns with prior work indicating that text-based private or sensitive information detected in photos such as names, addresses, or ID numbers, would be useful for blind people to decide whether to share them~\cite{zhang2023imageally}. P7 also suggested applying text extraction but this time to search for objects by their product labels to identify incorrectly-captioned photos, often in cases where objects are nearly identical by touch (\eg, different bags of snacks): \textit{``Let's say I picked up Cheetos, but accidentally called it Fritos...A search feature could tell me that because it's coming back with text of Cheetos, but I searched for Fritos. So you could actually get quite a bit of power with the search feature, particularly paired with text.''}
  
\noindent\textbf{Descriptors.} Some participants (P5, P7, P9) suggested new descriptors that can provide contextual information about the background, objects, and evaluation of photos. Specifically, they referred to \textit{``indoors or outdoors''} or \textit{``rug''} in the background (P5); \textit{``dimensions''} or \textit{``color''} of the object detected (P7); and \textit{``success rate''} for each (testing) photo as part of knowing the degree of recognition performance (P9). There were also suggestions to improve existing descriptors by specifying the number of objects detected in a photo as opposed to \textit{`more than one object detected'} (P3, P5), as well as detecting tattoos in addition to \textit{`face presence'} as identifiable information (P3).
    
When prompted, half of the participants (P1, P3, P4, P5, P7) expressed interest in adding future verbosity controls over descriptors, including activation and deactivation. One reason was \textit{``Because there might be stuff that people just don't care about. And that may vary slightly from person to person''} (P4). To enable such controls, participants suggested toggling certain descriptors via \textit{``checkboxes in settings''} (P1) or clicking on an image to access \textit{``more detailed descriptions''} such as object dimensions and colors (P7). 

\subsection{What Were Their Views on Mutliuser Access}

We unraveled blind participants' attitudes toward asking for sighted help in deciding whether to share their study data. Recognizing the limitations of the automatically generated image descriptors, some participants (P1, P3, P4, P6) acknowledged that they would resort to sighted help when needed. Yet, among all participants, only one (P4) sought sighted help from a family member to review their study data; most interacted with their study data independently prior to the co-design session. This aligns with prior work~\cite{jung2022expectations},  where blind people avoided asking sighted friends or family for detailed photo description to not \textit{``burden''} them. A few participants (P7, P10) expressed stronger concerns about relying on sighted help as it could hinder their independence. P7 referred to remote sighted help services to express similar concerns: \textit{``I think they have set us back into independence not forward...I have watched people where they could have figured something out themselves, call the services instead. And it has actually hurt us a great deal. You know, greatest harm, best intentions, all that.''} More so, finding `trusted' sighted help was reported to be a challenge (P1, P6, P7, P10), making the use of automated descriptors an ideal solution. 

Hence, when specifically asked about potential design features that allow access to `peripheral users' such as sighted help of their choice, half of our participants (P1, P5, P7, P8, P9) found it unnecessary or left concerns such as the reliance on someone's availability. P1 said, \textit{``That would mean having to have someone available when the sharing needed to take place, and not everyone's going to.''} Even among those who expressed it as a nice feature to share study data with sighted help of their choice, some of them (P3, P6, 10) articulated the difficulty of finding one or the desire to limit sighted help. For example, P3 said, \textit{``Even though I don't know 100\% of what was in all the photos. Except for that, as long as I can exclude all those with a face, I'm perfectly happy to do it [review the data] myself.''} 

While automated descriptors could be an ideal solution for independence and even offer speed with near-real time data access, there is inherently a tradeoff with accuracy for data interpretation. When exploring this tradeoff with our participants, we found that opinions are divided. Few (P1, P3) wanted to prioritize accuracy even if that meant waiting for manually annotated descriptors, others (P4, P8, P9) opted for speed, though some (P5, P6, P7, P10) recognized that a balance between the two is best.

Accuracy was deemed critical for making an informed decision, as specifically mentioned by P1 and P3. For example, P3 said, \textit{``I would choose accurately because if I don't have accurate information, I don't believe I can make the best decision.''} On the other hand, the common rationale among others who did not prioritize accuracy, was mainly associated with practicality given a large dataset, recognizing higher labor for manual annotation (P8, P9). There was also a negative point raised by P7 for \textit{``random people looking at it''} if image descriptions were to be manually created and perhaps outsourced, a common strategy for image captioning and annotation~\cite{bigham2010vizwiz,simons2020hope}: \textit{``Computers only do what computers are programmed to do. People, well, people aren't nearly so predictable.''}

In favor of speed, some participants (P4, P6, P9) highlighted the importance of maintaining continuity within the study context and receiving immediate feedback to address any concerns in deciding about their study data, suggesting \textit{``a quick turnaround time''} (P9). This perspective was similarly observed in prior work~\cite{morrison2023understanding} where immediate feedback was deemed critical for personalized accessibility. However, to a few participants (P1, P10), this meant \textit{``haste makes waste,''} where rushing through the process can impede their ability to make informed decisions. It was common for participants to opt for a balanced approach, as noted by P6: \textit{``I like things to be quick, because I won't remember what I did if it drags on too long, but by the same token, it would be nice to have a little more accuracy.''}

To strike a balance between accuracy and speed, some participants (P4, P5, P9) proposed a few strategies. One suggestion involved employing other forms of technology like text recognition to complement photo captioning methods (P7). P5 also suggested prioritizing the sharing of ``quality'' over ``quantity'' to reduce the volume of photos that would be reviewed and shared (P5). Additionally, P4 proposed integrating descriptors during data collection, enabling photo retakes and real-time resolution of any concerns---the original use of image descriptors in Hong \etal~\cite{hong2022blind}.

\section{Discussion}
The discourse around the ethical and inclusive development of AI has highlighted the benefits and responsibility of engaging blind people in contributing to datasets for AI innovation~\cite{kacorri2017teachable,theodorou2021disability, kamikubo2021sharing,sharma2023disability}. Extending such prior work on ``disability-first datasets,’’ we explored how to facilitate informed consent and meaningful data control for blind people when it comes to sharing their data via a public AI dataset. In this paper, we describe our data stewarding process involving 10 blind participants. They made decisions about their study data (\ie, photos of objects and labels) using a novel interface, AccessShare, to aid in forming and communicating their choices. Here, we further discuss how key insights from our study can inform future participatory data stewardship efforts as well as the implementation of systems supporting data access and control.

\subsection{Consent and Choices}

One challenge in our data stewarding with blind participants is that study data captured in their homes can contain unwanted or unintended elements that require further inspection and control mechanisms. Not to mention, this has been relevant to the broader topic of concern across the HCI and privacy literature~\cite{feng2021design,im2023less,gray2021dark}, and is growing widely with the rising trend of AI and data-driven innovation~\cite{ai2024model, hernandez2024secure}. Our work is situated in the context of understanding meaningful data controls for blind people, while considering the added challenge of inaccessibility of image data for visual inspection. This has previously left the researchers/data stewards to validate and decide on blind people's behalf which data are `safe' for public sharing~\cite{theodorou2021disability,sharma2023disability} after obtaining a blank consent early on, often prior to the data collection. 

Our findings showed that most blind participants preferred verbal consent during study sessions after data collection. Despite requests via email to decide on contributing their anonymized photos and labels to a public AI dataset, all but two participants did not respond. This trend also appeared among the three participants who completed the study but missed the co-design session as they did not respond to the final consent request. 

Although there may be various reasons for this lack of response, we observed that it is partly due to a need for a more interactive approach to consent. Blind participants highlighted the impact of engaging in dialogue with researchers, allowing them to discuss potentially concerning elements in their study data. Moreover, these discussions served beyond sighted help. Participants were encouraged to discern their decisions and concerns, as specifically mentioned by P1: \textit{``It has made me think a good deal about what would make me want to share and what wouldn't''} helping to \textit{``solidify''} their decision (not to share). Some also highlighted other learning outcomes that influenced their decision, by understanding the researcher's intent and exploring options that they could make, such as modifying their consent to share partially which P3 indicated as \textit{``I didn't know it was possible.''} 

Overall, the interactive approach to informed consent invites us to explore how it can be effectively operationalized. Beyond the practical challenges, what is the role of researchers in this interactive context? Do we need to account for power imbalances between participants and researchers~\cite{anderson2011national} in this process? In our study, participants that were initially inclined to share became more conservative after interacting with the data stewards (our team). Perhaps, this could be explained by the long collaborative relationship between our team and the blind community. However, we can see how this effect could go the other way, with the possibility of persuasion or coercion by data stewards. A recent example, though in a different context, coming from the ACCEPT-AI framework on pediatric data~\cite{muralidharan2023recommendations} could be used to discuss further on implementing interactive consent and safeguarding against potential harm. Authors argue about the importance of communication with the data contributors around risks, benefits, and \textit{alternatives}. Our study strengthen these perspectives and provides empirical findings that blind people shape their decisions when presented with alternatives or the `opportunity to choose'~\cite{walker2022value}. We recommend establishing and communicating the alternatives with potential data contributors; these are currently lacking in the AI research space~\cite{greene2024avoiding}.

\subsection{Implications for Data Sharing Platforms}

Insights from our study show us different expectations and needs among blind people for accessing their study data. For some participants, a platform like AccessShare was considered a complementary tool for their decision making, such as for communicating with the researcher to address any concerning elements and make granular decisions. Meanwhile, for one participant (P4) the complementary nature extended to making shared decisions with a family member.  In this case, communication with the researcher was not deemed critical. As such, we see metrics on `independence' and `interdependence'~\cite{reindal1999independence} affecting the effectiveness of a technology intervention for accessible data access and inspection. Below, we discuss design considerations informed from our findings.

\textbf{Negotiating independence and support.} Participants sought a delicate balance between self-reliance and seeking sighted help. They expressed concerns about relying on available and trusted sighted help beyond the data stewards when discussing the need for assistance with study data or a potential AccessShare feature for connecting with other sighted help. This indicates that narrowing the communication platform between data contributors and data stewards as part of data access and sharing could be helpful.  As suggested by a participant in our study (P8), it could incorporate a commenting feature for each photo to address any uncertainties or concerns. We already see some communication components being integrated into the work of Theodorou \etal~\cite{theodorou2021disability}, though the communication is initiated by the researcher \ie notify blind users of data validation results.  Having a dedicated person for data collection and management might be critical~\cite{jo2020lessons}, as Theodorou \etal eventually hired a contract researcher dedicated to the task~\cite{theodorou2021disability}. We would expect a similar challenge for addressing real-time feedback for this communication feature imagined by P8. In the long run, this might not be practical. While data-related clarification and communication could be complemented by AI-supported chat functionalities~\cite{xiao2023inform,allen2024consent}, the error-prone nature of this technology imposes its own limitations and risks.

\textbf{Enhancing data descriptors and inspection methods.} Probing with AccessShare was effective for ideating future data descriptors and inspection with blind participants. Some participants suggested integrating other AI-infused capabilities such as alt-text captioning~\cite{wu2017automatic, gurari2020captioning} or text recognition~\cite{deshpande2016real,neat2019scene}. These improvements could serve not only to enhance the photo descriptions but also to support their need to inspect for potential concerns. For instance, users can search the collection of photos based on text found on clothing (inspired by P3's feedback) or correct image classification errors by identifying any text on the object (inspired by P7's feedback). We also received suggestions for image descriptors to include contextual details about the background or the objects in the photos. Generative AI may seem like the obvious solution here for more comprehensive and nuanced descriptions (\eg~\cite{srinivasan2023automate}). Yet, it is a bit of a chicken and egg problem: \textit{blind people relying on AI to inspect their photos to help close the AI performance gap on photos taken by blind people.} Recent work by Massiceti\etal~\cite{massiceti2021orbit} shows that few-shot learning with a small number of photos can mitigate some of the quality-of-service disparities in current models but has not been yet examined for this scenario. Another design consideration is the need for verbosity controls. Participants recommended checkboxes and settings or hierarchical approaches for obtaining detailed descriptions. We see this as even more critical in the context of generative AI, where descriptions tend to be long and verbose, thus, impractical for a large number of photos.

\textbf{Considering multiuser data access.} As P4 indicated their joint decision-making approach with a sighted family member using AccessShare, we highlight considerations for including bystanders in the discussions regarding shared data access. This includes acknowledging the negotiation process in interactions between blind and sighted users, where blind users grant data access and control to sighted users in exchange for their time and insights. Conflicts of interest may also arise between them~\cite{lee2020pedestrian}. Considering these tensions, perhaps we ought to revisit the concept of consent as a shared multi-user decision. This has been widely discussed within accessibility (\eg, in the context of blind people using camera-based assistive applications~\cite{akter2022shared,ahmed2018up}) and beyond~\cite{mashhadi2014human,ohagan2023privacy}; however, Akter \etal has called for more exploration to understand how to involve both blind users and bystanders in the decision-making process~\cite{akter2022shared}. One component is addressing power dynamics that are increasingly concerning for technology-enabled abuse in co-occupant relationships~\cite{geeng2019who}. In our study, P5 specifically articulated potential misuse when granting different users access to inspect data: \textit{``Hopefully, the person who is reading it [inspecting the data] is not the type of person that will take advantage of that. Knowing that, oh, there's sensitive data in here. And then people are nosy.''} Thus, when exploring multiuser data access, power and control would need to be considered carefully to design varying levels of agency among users.

\subsection{Limitations and Future Challenges}
We presented a real-world data stewarding scenario designed to foster the involvement of blind participants in discussions around consent, data inspection, and control. Yet, there are limitations to the approach.
Our observations come from a small \textbf{sample}, even though $N=10$ meets the local standards for sample size in HCI~\cite{caine2016local}. 
While the sample is diverse in terms of age and balanced between women and men, it tends to have higher education backgrounds. Our study was remote, yet
participants were recruited from a relatively small area in the US near the authors’ institution. 
Some participants were recruited from our lab's mailing list, potentially familiar with the research team and with the concept of consent. Thus, caution is needed when generalizing the consent choices and preferences discussed in this paper to the broader blind community or new groups, as various factors, including trust and mutual respect~\cite{participants2013consent}, may impact them. 

Our approach could be replicated by researchers who are interested in gaining insights from the blind community in a different data collection context or studying approaches for involving another community of interest in data stewardship. To support such endeavors, we have shared the questionnaire used in our co-design study\footnote{Available at \textbf{\textit{\url{https://go.umd.edu/accessshare_questionnaire}}}}. Yet, we see how our method in its current form is limited as a practical tool for data stewards interested in obtaining informed consents from blind participants in their data collection. The process was \textbf{lengthy}, and scheduling multiple individual meetings with participants could be \textbf{prohibitive}, especially with the larger sample size. Based on the findings from our study, data stewards could choose to obtain consent via written or verbal means during (i) the interview, (ii) the co-design, (iii) or between the two sessions. However, when interpreting our findings, we would caution the possibility of an \textbf{order effect} on these options presented to participants for consent. While we observed their choices, it does not mean we identified the optimal option. It is most likely a combination that requires further exploration. 

One challenge we foresee from the absence of follow-ups, relates to the \textbf{control of participants' study data over time}. We did not receive any responses from the participants after the study to revoke their consent. Although implementing mechanisms for ongoing reassessment and consent could be important, there are practical challenges in answering: \textit{Who tracks consent over time? Does consent ever expire?} These anticipated questions across a dataset life cycle are worth considering~\cite{pushkarna2022data}, as failure to address them can lead to adverse impacts~\cite{longpre2024consent}. What's more challenging with consent in long-term engagement would be to rethink the way we approach research. For example, we would question our interactions with the research ethics committee when integrating dynamic consent practices---which involve, \textit{``bidirectional, continuous, and interactive communication between researchers and participants''}~\cite{lee2022toward}. Open-endedness could complicate the process and challenge researchers to constantly adjust consent protocols to different circumstances.

Last, we want to highlight observations and findings that were potentially influenced by our study and probe design. The first relates to PII in the photos taken by our participants using the smartglasses. We believe that \textbf{our Zoom setup during the data collection inflated the cases of ``object [photo of face].''} Profile pictures or videos, primarily of the experimenters but occasionally of the participants, were displayed on the laptop screen. Although unintended, it surfaced a common privacy issue when creating datasets containing image data sourced from blind people~\cite{gurari2019vizwiz}.

The second observation concerns the data descriptors in the AccessShare probe, initially designed to support blind users in machine teaching~\cite{hong2022blind}. Repurposing them for data sharing \textbf{misled many participants to see AccessShare as a reflection tool to improve the smartglasses application}, thinking that removing `bad quality' data (\eg, photos with cropped objects) from AccessShare could help. In hindsight, this is not a surprise. Participants perceived the original intent behind the descriptors. But this also surfaced an interesting design perspective: a dual-purpose system for data inspection and sharing, one that is bidirectional, enabling iterative training contributions~\cite{wang2023ai}. The challenge lies in balancing fewer quality issue photos for training versus more real-world quality issue photos for sharing. This involves addressing the value gaps between data stewards and data contributors, with many blind participants showing hesitancy to share `bad' photos, described as \textit{``tainted''} (P1). Co-developing these approaches and resources with blind people can be crucial. 

\section{Conclusion}
In this paper, we illustrated the active participation of 10 blind participants in our real-world data stewarding process, from collection to sharing of their study data. Within this process, we engineered a design probe, a novel data access interface called AccessShare, to examine their interactions with the data as they shaped their consent and to co-design accessible data sharing systems. A key aspect of our findings highlighted the impact of an interactive consent approach, which was further supported by the complementary nature of AccessShare in facilitating communication between data stewards and blind data contributors. Additionally, our study revealed design opportunities for data inspection and control functionalities. Yet, the participants' contrasting viewpoints on independence and interdependence in decision making raised nuanced considerations for designing multiuser data access to account for varying levels of agency among users. While the scope of this paper is to explore informed data sharing consents and meaningful data control mechanisms with blind people, our future work will be to make the dataset available to the broader research community. Moving forward, we plan to apply the insights gained from the study (including our analysis from PII annotations) and integrate participant consent and feedback into our data sharing practices.

\begin{acks}
We thank the blind participants in our study as well
as the anonymous reviewers for their comments that further strengthened this paper. Special thanks to Jessica Vitak for her feedback on our preliminary analysis, and the research advisory council from The National Federation of the Blind for their assistance with participant recruitment. This work is supported by National Institute on Disability, Independent Living, and Rehabilitation Research (NIDILRR), ACL, HHS (grant \#90REGE0008). Hernisa Kacorri was additionally supported from NIDILRR grant \#90REGE0024.

\end{acks}

\bibliographystyle{ACM-Reference-Format}
\bibliography{main}
\end{document}